\documentclass[aip,apl,amsmath,amssymb,reprint]{revtex4-1}
\usepackage{graphicx}
\usepackage{subfig}
\usepackage[version=3]{mhchem} % Formula subscripts using \ce{}
\usepackage{dcolumn}% Align table columns on decimal point
\newcolumntype{d}[1]{D{.}{\cdot}{#1} }
\usepackage{bm}% bold math
\usepackage{xcolor,soul}

\usepackage{siunitx}% JMF addition - once a physicist, always a physicist
\begin{document}
%\title{Can hybrid halide perovskites form an intrinsic intermediate band solar cell?}
% Maybe better with no question?
\title{A photon ratchet route to high-efficiency hybrid halide perovskite intermediate band solar cells }

\author{Jarvist M. Frost}
\affiliation{Department of Materials, Imperial College London, Exhibition Road, London SW7 2AZ, UK}
\email[Electronic mail:]{jarvist.frost@imperial.ac.uk}
%\affiliation{Centre for Sustainable Chemical Technologies and Department of Chemistry, University of Bath, Claverton Down, Bath BA2 7AY, UK}

\author{Pooya Azarhoosh}
\affiliation{Department of Physics, Kings College London, London WC2R 2LS, UK}

\author{Scott McKechnie}
\affiliation{Department of Physics, Kings College London, London WC2R 2LS, UK}

\author{Mark van Schilfgaarde}
\affiliation{Department of Physics, Kings College London, London WC2R 2LS, UK}

\author{Aron Walsh}
%\email[Electronic mail:]{a.walsh@bath.ac.uk}
\affiliation{Department of Materials, Imperial College London, Exhibition Road, London SW7 2AZ, UK}
%\affiliation{Centre for Sustainable Chemical Technologies and Department of Chemistry, University of Bath, Claverton Down, Bath BA2 7AY, UK}
\affiliation{Department of Materials Science and Engineering, Yonsei University, Seoul 120-749, Korea}


\begin{abstract}
    %
    % 100 word limit for APL (strict on submission)- reached now! 
    %
    %Are hybrid halide perovskites intermediate band solar cells?  
    %And if not, can we design a material that is?
    %...
    The spin-split indirect bandgap in hybrid-halide perovskites
    provides a momentum-space realisation of a photon-ratchet intermediate band.
    Excited electrons thermalise to recombination-protected
    Rashba pockets offset in momentum space, building up the charge
    density to have sufficient flux to the higher lying conduction band. 
    This effect could be used to form an intrinsic intermediate band solar cell 
    with efficiencies beyond the Shockley-Queisser limit
    if a selective low-electron affinity contact can be made to the higher conduction state.
    This concept is supported by analysis of the many-body electronic structure. 
    Production of above-bandgap voltages under illumination
    would affirm the physical mechanism proposed here. 
\end{abstract}

%\pacs{88.40.-j, 71.20.Nr, 72.40.+w, 61.66.Fn}
% 71.20.Nr 	Semiconductor compounds 
% 72.40.+w 	Photoconduction and photovoltaic effects
% 61.66.Fn 	Inorganic compounds 
% 88.40.jn 	Thin film Cu-based I-III-VI2 solar cells
% 88.40.-j 	Solar energy

\maketitle 

%\textbf{Introduction}

The Intermediate Band Solar Cell (IBSC)\cite{Luque1997,IBSC-Review2015} is a photovoltaic
device architecture that offers the potential to significantly exceed the
Shockley-Queisser (SQ) power conversion efficiency limit.
By introducing an additional intermediate band, there are three excitation
processes at different wavelengths that subdivide the solar spectrum. 
A working realisation has so-far been prevented by the fact that any state
introduced within the gap also provides a route for recombination.
IBSC cells considered to date create an intermediate band by doping
the crystal with defects, or by creating a quantum-dot superstructure.  
The high emission rate competes strongly with the absorption to the conduction band.

A necessary condition for an IBSC is that independent quasi-Fermi levels can be
maintained in each of the three bands. 
This requires that electrons cannot transport between the bands, either
directly (phonon mediated) or via a continuum of electronic states in the contacts. 

If the device is to exceed the SQ limit, the charge density in the
intermediate state must build sufficiently for balanced absorption 
to the conduction band.  A long-lived intermediate state has been
identified as a key requirement for a working IBSC.  This is
challenging with a single intermediate band, since spontaneous
emission back down to the valence band also increases with the
occupation of the intermediate band.  
Recently, the photon ratchet mechanism was proposed\cite{PhotonRatchet2012},
where the photoexcited intermediate band thermalises into another state,
with reduced recombination rate to the ground state.  
This process significantly increases the lifetime of the intermediate state
and provides a promising route for realising high-efficiency IBSC.

\begin{figure}
    \includegraphics[width=\columnwidth]{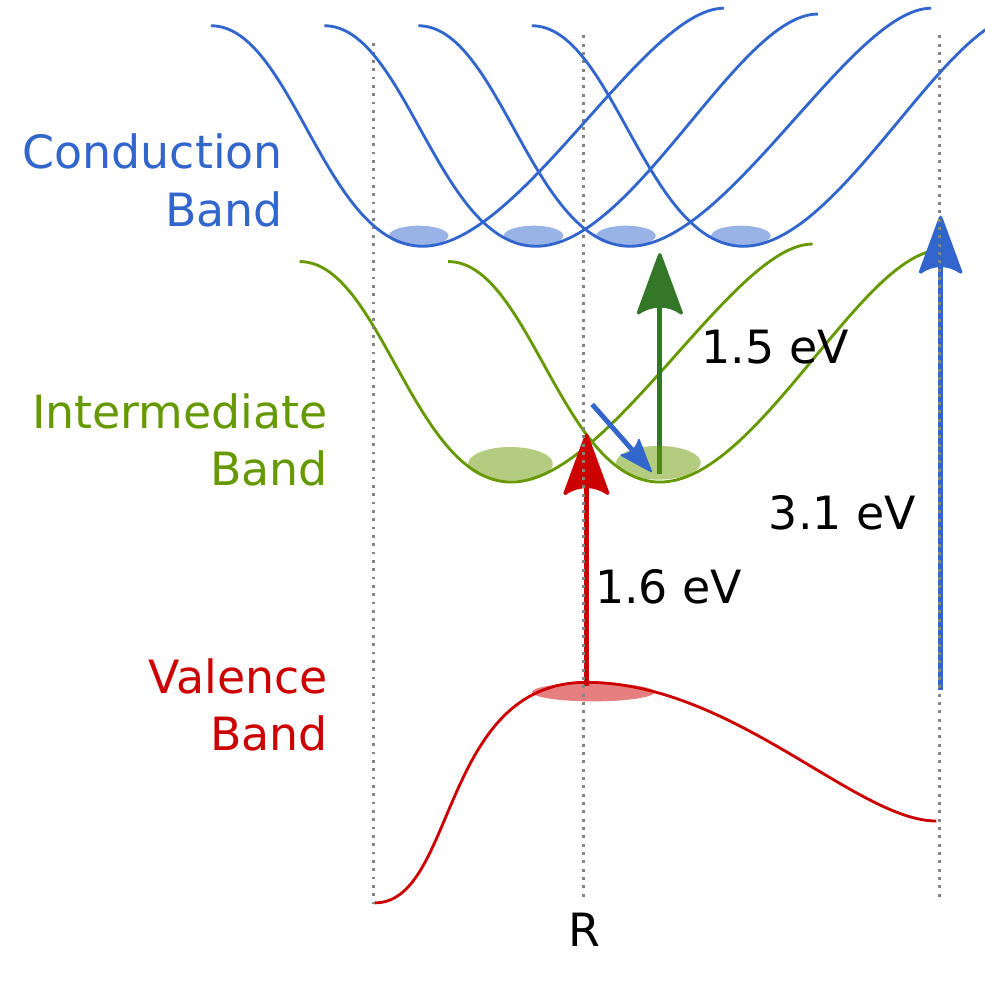}
    \caption{\label{MAPI-Ratchet-schematic}A schematic of the proposed hybrid halide
    perovskite spin-split indirect-gap photon-ratchet. 3.1\,eV excitation
    directly pumps to the higher lying conduction band. 
    1.6\,eV excitation pumps to the spin-split indirect-gap intermediate band,
      where fast thermalisation relocates carriers slightly off the high
      symmetry location. Low hole density of states in the valence band means
      that direct recombination is severely reduced. 
      Charges in this state can be excited to the second conduction band by
      a ~1.5 eV excitation. 
      Therefore, the direct 3.1 eV excitation and the two-step photon-ratchet
      proceed simultaneously, and energy can be extracted from the higher lying
      conduction band.
    }
\end{figure}

%\textbf{Spin-split indirect-gap}
In this Letter, we propose that the spin-split indirect-gap\cite{Bcoeff2016} in
methylammonium lead halide perovskite (\ce{CH3NH3PbI3}) forms an intermediate band.  
A photon-ratchet effect is provided by carriers excited into this band,
relaxing in momentum space to the Rashba pocket (see Figure
\ref{MAPI-Ratchet-schematic}).  
This mechanism is intrinsically present in the bulk material. 

Electron-hole recombination to the  valence band (VB) from both the intermediate band (IB) and
conduction band (CB) is reduced by the slightly indirect gap originating from the
mutually-orthogonal Rashba extremal points. 
%Theramlised (Fermi-Dirac) density of states have a very low dual-density due to
%the momentum offset.
Both the middle bands and the conduction bands are spin split
due to spin-orbit coupling, but along
different axes; thus none of the bands have extremal points at the same
$k$ in momentum space.  
A quasi-equilibrium distribution of electrons give rise to a
low joint density-of-states for vertical transitions. 
%
%The next higher-lying conduction band about 1.5\,eV above the usual conduction
%band, plays the role of the conduction band.  
%
%Scott comment: we discussed this and the Rashba splitting shouldn't changes
%the recombination from CB2 to CB1 as CB1 is a mostly empty band. The deviation
%from parallel bands will reduce the recombination rate but this will be a small
%contribution. The absorption rate from CB1 to CB2 will be slightly increased by
%the splitting as the electron density will not build up at the vertical
%transition point.)) 
%
We invoked a similar mechanism\cite{Bcoeff2016} 
for conduction to valence band transitions
in order to explain and quantify
the long charge-generated carrier recombination times in lead halide
perovskites with standard device architectures.

The excitation from the VB to IB is
at 1.6\,eV, while the direct VB to CB excitation at 3.1\,eV.\cite{Ellipso2016}
These states are optically bright.  
The photon ratchet from the IB to the CB would operate at
1.6\,eV.
These energies are not well matched to subdivide the solar spectrum,
but they are accessible in the laboratory.  
To exceed the SQ limit, a smaller bandgap material is required. 
However, we propose lead halide perovskites as a well understood test
system in which the physics can be studied, to leverage expertise in
making high-quality materials and devices.
A working IBSC should also be an upconverter when operating at open circuit (or
equivalently, with no contacts). 
This will require a pure material where non-radiative recombination pathways to
be sufficiently suppressed to observe light output. 

The higher lying conduction band provides a high dispersion (low effective
mass) band from which to extract the charge carriers. 
%The same slightly-indirect gap exists between the Rashba pocket photon-ratchet
%and the higher conduction band---very low recombination will occur via the
%photon-ratchet while the solar cell is in operation (extracting charge and so
%reducing charge density in the conduction and valence bands). 
The only change in device architecture required is a low work-function
electron-accepting contact (such as Ba, Ca, LiF, or fulleroid adduct), to
selectively collect from the higher energy CB. 
With suitable contacts, the material should thus be able to generate an
anomalous photovoltage, above circa. \SI{3}{\electronvolt}) when
pumped with incident light of 1.6\,eV.

\textit{Qualitative model:} 
There are two necessary conditions for a working IBSC:
(i) the CB and IB should develop independent quasi-Fermi levels;
(ii) the CB must be electrically contacted independent of the IB. 
We can assess whether these conditions are achievable by inspecting the 
 electronic band structure. 

Band structures are usually presented as cross-sections of the
Brillouin zone, with the path chosen to follow an irreducible representation
of the underlying crystal symmetry. 
From the Bloch theorem, these high symmetry lines form the extrema and turning
points in the electronic structure, and so fully characterise the band
functions. 
In hybrid halide perovskites, the presence of a molecule (and large dynamical flexing of the octahedral cage) breaks the local
$O_h$ symmetry of the underlying pseudo-cubic lattice. 
Practically, the
spin-orbit coupling (due to atoms with large nuclear charges) interacting with
local crystal fields moves band extrema to off-symmetry locations. 
Amongst other effects, this results in the splitting of the lower
conduction-band to give a spin-split indirect-gap. 
To guarantee that we are exploring all possible routes for recombination and
thermalisation, we must integrate over the Brillouin Zone.
By considering the reciprocal-space resolved difference in energy eigenstates,
we can demonstrate whether the two conditions for an IBSC are met.

We use the QS\textit{GW} electronic structure method as implemented in the Questaal
codes\cite{Questaal}, improving on our prior work\cite{Brivio2014} by using
a larger and more converged basis on the same pseudo-cubic structure.
It has been established\cite{Chantis06a,Krich07} that QS\emph{GW} yields very good
Dresselhaus splitting in semiconductors, in contrast to the LDA.
For  convenience, the spin-orbit coupling was added only as a post-processing step: 
if it is included in the self energy the gap is reduced about 0.1\,eV\cite{Brivio2014}).
These codes allow for the calculation of a self-consistent self-energy on
a regular \textit{k}-point mesh, which can then be used for
calculation at an arbitrary \textit{k} point. 
This allows for sufficient finesse to locate and characterise the slightly
off-symmetry extrema. 
Thus we consider this method to be of the highest quality available for
broken-symmetry relativistic-ion semiconductors such as \ce{CH3NH3PbI3}.

\begin{figure}
    \includegraphics[width=\columnwidth]{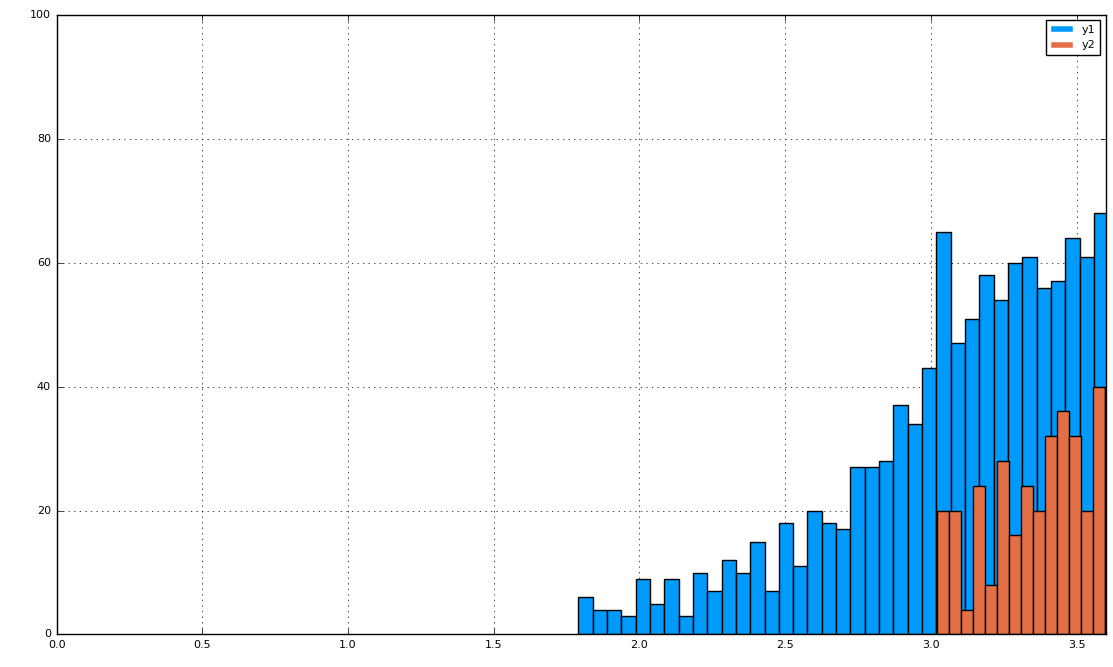}
    \caption{\label{mapi-jdos}
    Joint densities of states (vertical transitions only) from a QSGW
    calculation including spin orbit coupling. 
    Blue shows the CB to IB transitions (starting at 1.6\,eV), orange shows the
    VB to CB transitions (starting at 3.1\,eV). 
    The photon ratchet is at 1.5\,eV, but the joint density of states is
    strongly dependent of out of equilibrium population in that band. 
    }
\end{figure}

We find no connection between bands in an energy window around the photo
excited charge carriers, but at very high energies above the Fermi level, the
bands do come together. 
This suggests that distinct quasi-Fermi levels are experimentally realisable. 
A consideration of the joint densities of state indicate that both the VB to IB
and VB to CB photoexcitation channels will be simultaneously operational
(Figure \ref{mapi-jdos}).

The band-edge electronic structure for hybrid perovskites is unique:
there is a dispersive upper valence band (formed of I $p$ orbitals),
an intermediate band (mainly hybridised Pb $p$ orbital), and the upper conduction
band (two-fold degenerate, mainly hybridised Pb $p$ orbitals). 

These extrema are centered around the high symmetry $R$ point in the
Brillouin zone. 

The spin-orbit coupling (due to large nuclear charge on Pb and I, and
associated relativistic term in the Hamiltonian) interacts with the local
crystal field to split the spin degeneracy, projecting differing spin electrons
to antipodal locations around $R$.  
This effect leads to the spin-split indirect-gap. 
As the orbital character is different for all three bands, we find that the
reciprocal space projection is mutually orthogonal. 
Spin-split indirect-gaps exist between all VB, IB and CB. 

\begin{figure}
    \includegraphics[width=\columnwidth]{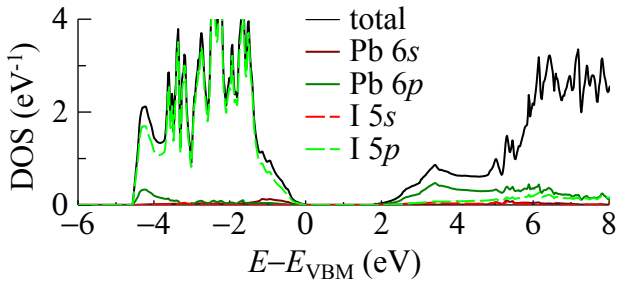}
    \caption{\label{mapi-pdos}
    Partial densities of state from a QSGW calculation including spin orbit
    coupling. 
    A Mulliken projection of the total DOS is made onto Pb 6$s$ and 6$p$, and 
    I 5$s$ and 5$p$.
    The valence band is shown to be almost pure I 5$p$, as would be expected from
    a tight-binding picture.
    However, the intermediate and conduction bands are not purely of Pb 6$p$
    character, as would be predicted by atomic orbital arguments. 
    This suggests that transitions between the intermediate and conduction band
    are not symmetry forbidden, as would be the case if they were both composed
    wholly of identical orbital angular momentum states. 
    A more complete understanding requires calculation of explicit
    dipole matrix-elements, which will require custom codes. 
    }
\end{figure}

% First table - Extrema positions in k-space
\begin{table}[h]
\centering
\begin{tabular}{llcccr}
\hline
    Band & $d$ & $k_x$ & $k_y$ & $k_z$ & $|R-k|$ \\ 
\hline
    VB & 2 & $R\pm0.0072$ & $R$ & $R\mp0.0233$ & $0.0244$ \\ 
    IB & 2 & $R\mp0.0220$ & $R$ & $R\pm0.0436$ & $0.0489$ \\
    CB & 4 & $R\mp0.0273$ & $R\mp0.0452$ & $R\pm0.0396$ & $0.0660$ \\ 
\hline
\end{tabular}
\caption
{\label{kmin}
    Extremal location of upper valence band (VB), intermediate band (IB), and lower conduction band (CB)
    with degeneracy (\textit{d}).
    Units are \si{\per\angstrom}, all quoted
    relative to the $R$ high symmetry location.
  %  (( JMF --- I think these are very interesting, as is the rotation of the
  %  +/- for the 4 fold degenerate structure. 
  %  I've also generated some 3D renders of the effective mass, with ellipsoids
  %  filling out the $W(k)=\frac{h^2}{2m^*}k^2$ ellipsoids. 
  % But I'm uncertain how to deal with the principle axes.
  %  Certainly it seems the minima are distributed around $\Gamma$-R.
}
\end{table}

% Rashba energy one
\begin{table}[h]
\centering
\begin{tabular}{lcccc}
\hline
    Band & $\Delta E_{Rashba}$ & $m_x^*$ & $m_y^*$ & $m_z^*$ \\
\hline
    VB &  $9.5$ & \num{-0.142} &  \num{-0.241} &  \num{-0.892}\\
    IB & $63.7$ & \num{1.250}  &  \num{0.153}  &  \num{0.128}\\
    CB & $19.5$ & \num{0.583}  &  \num{0.290}  &  \num{0.191}\\
\hline
\end{tabular}
\caption
{\label{rashba}
    Rashba energy depth (from extremal point to high-symmetry saddle location) in
   \si{\milli\electronvolt}
    and effective mass ($m^*$ in units of $m_e$) calculated around the Rashba extremal location for the three bands.
}
\end{table}

In the equivalent circuit of an IBSC, the VB to IB and IB to CB transition are
connected in series by the conservation of electron charge; forming 
equivalent circuit of a two-lead tandem solar cell.
Therefore an efficient device requires the current flux between the two
processes to be balanced. 
%In MAPI, the energy of these excitations is very close. 
This depends on the details of the band structure, including the bandgap, 
density of states, and the strength of the optical transitions (transition matrix elements). 
%
% AW - I cut a lot below - the following is a little raw, but trying to get more to the point
%
The critical factor for hybrid perovskites, which is not present in tetrahedral
semiconductors such as GaAs, is the reduced electron-hole recombination rates
due to Rashba splitting.
As discussed above, these will affect CB $\rightarrow$ VB and IB $\rightarrow$
VB transitions, due to a momentum offset between thermalised populations of
electrons and holes.
Our prior analysis\cite{Bcoeff2016} has shown that the direct radiative
recombination rate is reduced by a factor of  350X % fill me in
under one sun illumination.
This will ensure significant steady state population of electrons in the IB under working conditions.

In summary, we have proposed a novel route to
achieve intermediate band solar cells based on spin-split indirect-gap materials
such as hybrid perovskites.
According to the calculations presented herein, with suitable contacts
\ce{CH3NH3PbI3} should make a working device, with all transitions in the
visible, and accessible in the laboratory. 
In order 
to quantitatively predict whether a device with efficiencies beyond the SQ 
limit is achievable will require more detailed device
simulations that takes into account 
the unusual band physics, including the asymmetry in absorption and recombination
between the three band system (VB, IB and CB).  

Beyond \ce{CH3NH3PbI3}, in looking for spin-split materials to make high power conversion 
efficiency IBSC devices, we want to identify materials with a lower band-gap
than the SQ matched 1.5\,eV.  
Searches of novel materials for photovoltaics often discard such
low-gap materials. 
An exciting prospect of using a bulk, band structure engineered, material for
IBSC is the relatively low cost of making the devices, compared to engineering
real-space heterojunctions. 

\begin{figure}
    \includegraphics[width=\columnwidth]{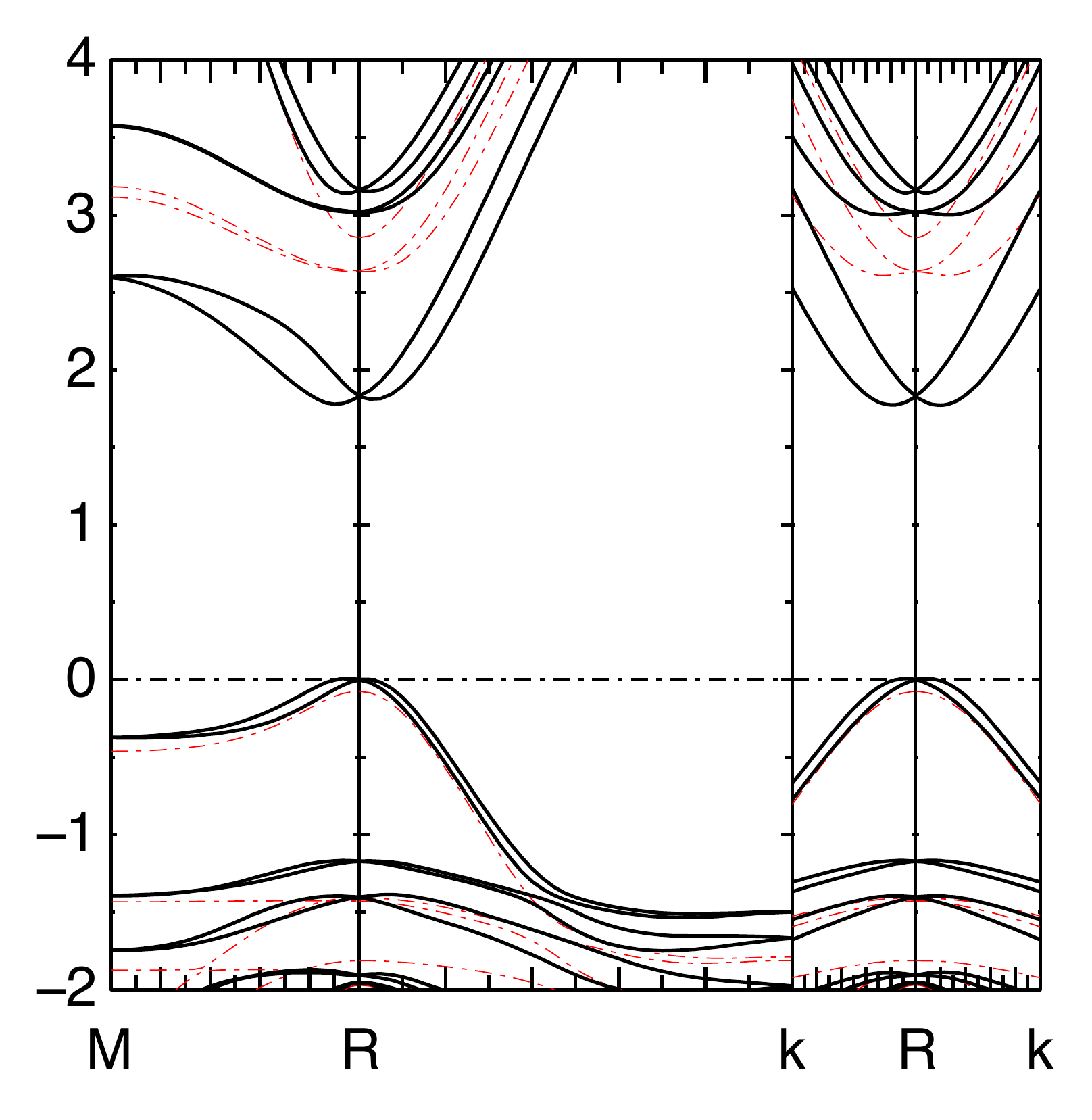}
    \caption{\label{MAPI-Bands}
Many-body electronic band structure of \ce{CH3NH3PbI3} from QS\textit{GW}.
    The Red-dashed line shows the band structure without a spin-orbit
    contribution, the three ${p}$ orbital conduction bands are near 
    3.0\,eV, split in a two-fold and singly degenerate level. . 
    Including spin-orbit (Black, full), the energies are significantly
    renormalised with the singly-occupied band shifting down to 1.6\,eV, and
    the spin-channels Rashba split into two (IB) and four (CB) mutually
    inequivalent minima.  
%    [Better caption from Scott to explain the right panel]
}
\end{figure}

\begin{figure}
    \includegraphics[width=\columnwidth]{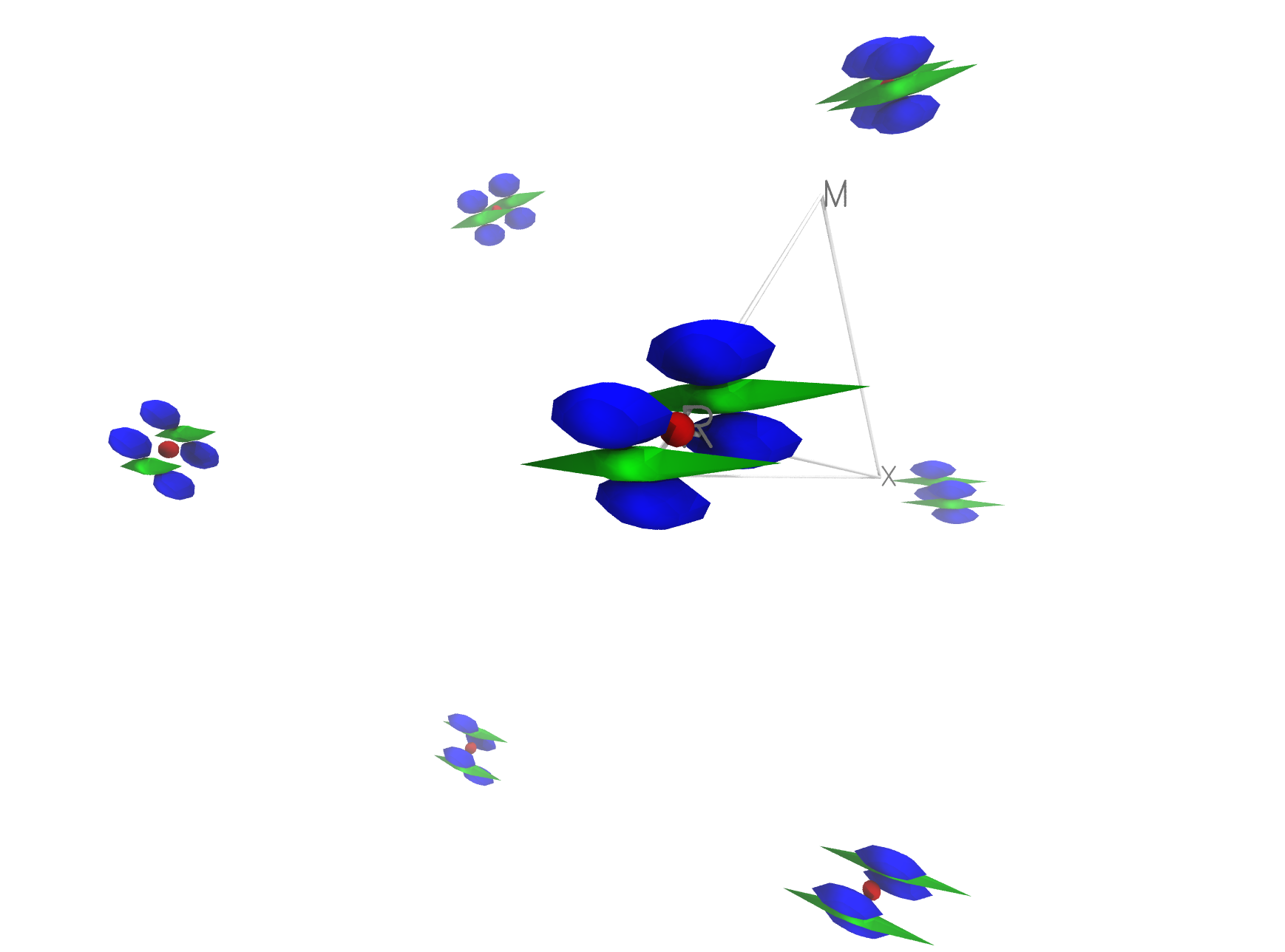}
    \caption{\label{Rashba-render}
    A 3D render of the Rashba pockets of the VB (red), IB (green) and CB
    (blue), located around the high symmetry R locations. 
    The ellipsoids are generated from the sampled effective mass tensors. 
    %and are integrated
    %out to an energy of XXX meV. 
    %Total volume of reciprocal space (and thus electron density) enclosed by
    %these volumes is YYY electrons / unit cell, ZZZ \si{\per\centi\metre\cubed}.
    %(( JMF --- actually, I think these are $m^{*}$ wide ellipsoids, rather than
    %$(m^{*})^{-1}$ as they should be.
    %Also, these are just the Cartesian orthogonal axes, rather than the
    %principel axes supplied by Scott. 
    %I need to figure out the unitary transformation for one onto t'other. )) 
    }
\end{figure}

\textbf{Acknowledgement}
We thank Jenny Nelson and N. J. Ekins-Daukes (NED), for stimulating discussion. 
We acknowledge membership of the UK's HPC Materials Chemistry Consortium, which is funded by EPSRC grant EP/F067496. 
%FIXME: This is probably old HPC MCC grant #? 
J.M.F. is funded by EPSRC Grant EP/K016288/1, the KCL group acknowledges support from EPSRC Grant EP/M009602/1.
A.W. acknowledges support from the Royal Society. 
The authors declare no competing financial interests. 

\bibliography{MAPI-Ratchet}

\end{document}